\documentclass[a4paper,11pt]{article}

\usepackage{amsmath}
\usepackage{amssymb}
\usepackage{amsfonts}
\usepackage{verbatim}
\usepackage{graphicx}
\usepackage{epstopdf}
\usepackage{caption}
\usepackage{enumitem}
\usepackage{color,xcolor}
\usepackage[pdfstartview=FitH,
            bookmarksnumbered=true,
            bookmarksopen=true,
            colorlinks=true,
            linkcolor=blue,
            citecolor=blue,
            urlcolor=blue,
            anchorcolor=blue
            ]{hyperref}

\newcommand{\wjp}[1]{\textcolor{black}{#1}}
\newcommand{\pz}[1]{\textcolor{black}{#1}}

\author{Pu Zhang $^1$ \footnote{Email: pzhang@binghamton.edu} \ and William J. Parnell $^2$ \footnote{Email:William.Parnell@manchester.ac.uk}}

\date{%
	$^1$\footnotesize{Department of Mechanical Engineering, Binghamton University, Binghamton, NY 13902, USA}\\%
	$^2$\footnotesize{School of Mathematics, University of Manchester, Manchester, M13 9PL, UK}\\[2ex]%
	\today
}

\title{Hyperelastic Antiplane Ground Cloaking}

\begin{document}

\maketitle

\begin{abstract}
Hyperelastic materials possess the appealing property that they may be employed as elastic wave manipulation devices and cloaks by imposing pre-deformation. They provide an alternative to microstructured metamaterials and can be used in a reconfigurable manner. Previous studies indicate that exact elastodynamic invariance to pre-deformation holds only for neo-Hookean solids in the antiplane wave scenario and the semi-linear material in the in-plane compressional/shear wave context. Furthermore, although ground cloaks have been considered in the acoustic context they have not yet been discussed for elastodynamics, either by employing microstructured cloaks or hyperelastic cloaks. This work therefore aims at exploring the possibility of employing a range of hyperelastic materials for use as antiplane ground cloaks (AGCs). The use of the popular incompressible Arruda-Boyce and Mooney-Rivlin nonlinear materials is explored. The scattering problem associated with the AGC is simulated via finite element analysis where the cloaked region is formed by an indentation of the surface. Results demonstrate that the neo-Hookean medium can be used to generate a perfect hyperelastic AGC as should be expected. Furthermore, although the AGC performance of the Mooney-Rivlin material is not particularly satisfactory, it is shown that the Arruda-Boyce medium is an excellent candidate material for this purpose.
\end{abstract}

\section{Introduction}

Early efforts to develop invisibility cloaks were initiated in the work of \cite{greenleaf2003nonuniqueness, pendry2006controlling} and \cite{leonhardt2006optical} by employing the method of transformation optics and shortly after this a variety of electromagnetic cloaks were designed, including those in free-space \cite{ruan2007ideal,schurig2006metamaterial,cai2007optical} and ground cloaks \wjp{(or carpet cloaks, both terms are used in the literature)} \cite{li2008hiding, liu2009broadband, ma2010three}. The notion underlying transformation methods \cite{leonhardt2009transformation} is to construct an inhomogeneous and frequently anisotropic material in the physical space that has been derived via a conformal mapping from a uniform virtual space, so that they have equivalent responses to observers. Transformation methods were soon extended to manipulate waves beyond the electromagnetic regime, including the application to acoustics \cite{norris2008acoustic, chen2010acoustic, popa2011experimental, zigoneanu2014three} and elastodynamics  \cite{diatta2014controlling,hu2011approximate,liu2015transformation, chang2012transformation,stenger2012experiments, climente2016analysis} based on the form-invariance of the wave equations under coordinate transformations. Unlike transformation optics or acoustics however transformation elastodynamics does not always guarantee the form-invariance of Navier's equations under generalized coordinate transformations \cite{milton2006cloaking,norris2011elastic, brun2009achieving, colquitt2014transformation}. This makes it more challenging to control elastic waves. A more general set of equations, associated with inhomogeneous elastic materials now known as the \textit{Willis equations} does guarantee invariance but with the added difficulty of more complex constitutive behaviour \cite{milton2006cloaking,norris2011elastic}. Recently symmetrised elastic media have been employed to create near-cloaks in the context of elastodynamics \cite{sklan2017elastic}. One of the exceptions however in the context of elastodynamics is the antiplane shear wave case, since the elastic displacement in this two-dimensional scalar problem is governed by Helmholtz' equation, similar to acoustics. Consequently, antiplane elastic cloaks have been designed successfully based on the transformation method \cite{liu2015transformation, wu2015manipulation}. Despite significant progress in the context of both theoretical and experimental aspects of antiplane cloaking, most of the aforementioned cloaks adopt materials with complex microstructures, which are generally difficult to design and fabricate. Furthermore although ground cloaks have been designed in the context of acoustics \cite{popa2011experimental, zigoneanu2014three} and there has been much recent discussion of cloaking surface waves \cite{khlopotin2015transformational, colombi2016transformation, colombi2017elastic}, no designs have yet been proposed for ground cloaks in the elastodynamic context.

An alternative mechanism to design elastic cloaks and also to manipulate the propagation of elastic waves more generally is based on the so-called theory of \textit{hyperelastic cloaking} \cite{norris2012hyperelastic, ParnellBookChapter2017}. This theory was first developed for antiplane shear waves \cite{parnell2012nonlinear, parnell2012employing} but shortly after was extended to coupled in-plane compressional/shear wave problems \cite{norris2012hyperelastic}. Recently, this theory has also been employed to design phononic crystals with band gaps that are invariant to pre-deformation \cite{zhang2017soft}. \pz{Instead of achieving conformal mapping by using microstructural units, hyperelastic cloaking theory utilizes the transformed elasticity tensor of a rubber-like material under pre-deformation, which avoids the design of complex microstructures \cite{norris2012hyperelastic, parnell2013antiplane}}. It has been shown that exact hyperelastic cloaks can only be achieved for special cases of hyperelastic materials, e.g.\ antiplane cloaking by using neo-Hookean solids and in-plane cloaking by using semi-linear materials under typical deformation modes \cite{norris2012hyperelastic}. This constraint has restricted the wide-spread application of hyperelastic cloaks for practical usage. In \cite{parnell2013antiplane} hyperelastic cloaking using neo-Hookean and Mooney-Rivlin materials was compared to layered microstructures for the case of a two-dimensional cylindrical cloak in free space.

Until now then, most elastic cloaks reported in the literature are of the free space type, while elastodynamic ground cloaks have not yet been explored. An antiplane cloak could be devised from microstructure by employing e.g.\ layered materials but here we devise a hyperelastic antiplane ground cloak. \wjp{It should be noted that hyperelastic metamaterials have several potential advantages over microstructured cloaks. These include tunability, frequency independence and the fact that the cloaked region can be made larger or smaller by imposing more or less deformation so that the cloak is not fixed once in-situ.} We aim to explore the possibility of using hyperelastic materials other than a neo-Hookean medium to design antiplane ground cloaks. The hypothesis is that some hyperelastic materials can still be used as approximate cloaks, albeit that there is still some difference between the cloaked and uncloaked wave responses. In order to achieve the objective of generating a hyperelastic antiplane ground cloak, we consider materials with neo-Hookean, Arruda-Boyce, and Mooney-Rivlin strain energy functions. The scattered wave fields for the cloaked and uncloaked problems are simulated by finite element analysis and are compared in order to evaluate the applicability of the corresponding hyperelastic materials.

The paper is organized in the following manner. The basic theory for hyperelastic ground cloak problems is introduced in \S \ref{sec:theory} with the emphasis being the antiplane elastic wave context. The finite element simulation model and associated detail will be introduced in \S \ref{sec:model} with results and discussion presented in \S \ref{sec:results}. Conclusions are given in \S \ref{sec:conclusion}.

\section{Hyperelastic Ground Cloak} \label{sec:theory}

\begin{figure}[htbp!]
	\centering
	\includegraphics[width=12cm]{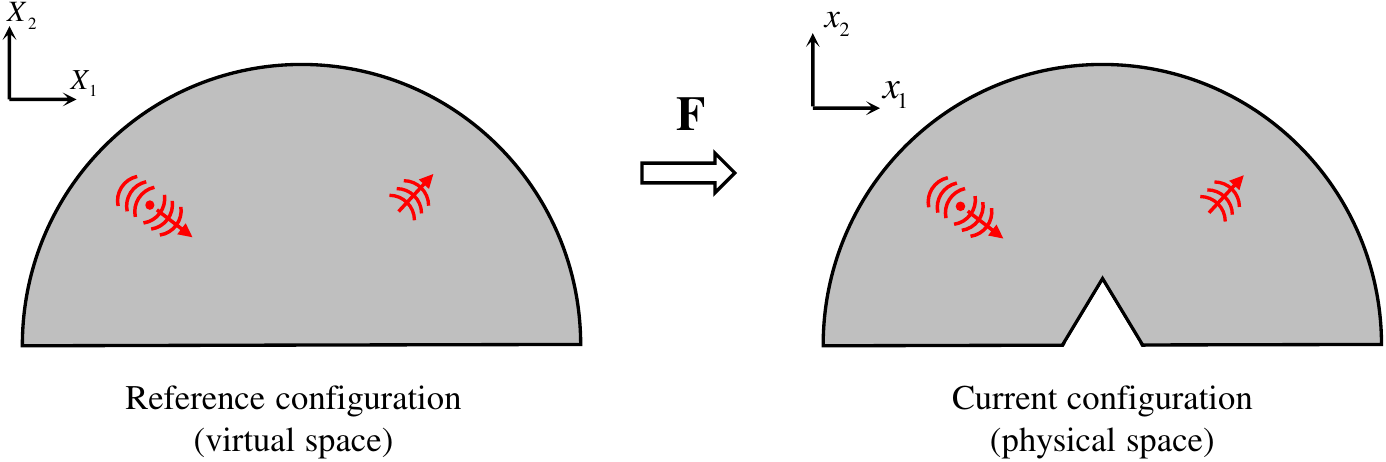}
	\caption{\label{fig:groundcloak} Schematic illustration of a hyperelastic ground cloak with respect to the \wjp{reference configuration $\mathcal{B}_r$ (left) and current configuration $\mathcal{B}$ (right) respectively}. The cloaked region is obtained by imposing a pre-deformation field $\mathbf{F}(\mathbf{X})$. }
\end{figure}

Hyperelastic cloaks retain the form-invariance of the \wjp{linear} elastic wave equations by imposing pre-deformation and by choosing specific strain energy functions. In this paper we illustrate the potential of using hyperelastic media for \textit{ground cloaking}. Here we show that hyperelastic materials in a pre-deformed state can be employed as ground cloaks in the antiplane wave scenario. The configuration is depicted in Fig.\ \ref{fig:groundcloak} as a half space with its free boundary initially residing along $X_2=0$ and with the space extended to infinity in $X_2>0$.  A wave source generates the incident antiplane wave field that impinges on the free surface. The cloaked region is created by imposing local deformation on surface of the half-space, depicted as a triangular indented region here but in practice it will take a more complex geometry depending upon the mechanism of indentation. Generally the presence of such an indentation of the free-surface would mean that waves would be scattered from it whether in a deformed state or otherwise, \textit{but we wish to choose a nonlinear material such that in the pre-deformed state, antiplane waves will scatter from the indentation as if the surface remains flat and traction free}. \wjp{This is the purpose of a ground cloak.}
\wjp{Based on continuum theory, a material point can either be described by the material coordinate $\mathbf{X}$ in the reference configuration $\mathcal{B}_r$ or by the spatial coordinate $\mathbf{x}$ in the current configuration $\mathcal{B}$}, which are related by the mapping \pz{ $\mathbf{F}(\mathbf{X}) = \partial \mathbf{x} / \partial \mathbf{X}$}, known as the deformation gradient. The wave responses are equivalent in the reference and current configurations as an analog to the virtual space and physical space in the transformation theory. Thus, a cloak is formed when the pre-deformation field $\mathbf{F}(\mathbf{X})$ does not affect the wave field propagating in the virtual space.  In \cite{parnell2012nonlinear,norris2012hyperelastic} an equivalence theory was derived whereby transformation elastodynamics was forced to be equivalent to the hyperelastic theory of `small-on-large' \cite{ogden2007incremental}, thereby giving a specification on a required strain energy function to force independence of the wave field from the deformation gradient.  Note that the boundary condition in the deformed configuration can potentially also have an important influence on the subsequent scattered field. Here, an in-plane indentation is imposed \wjp{and it transpires that, assuming the interface between the indenter and the deformed body is friction free, the imposed in-plane traction ensures that incremental antiplane waves see a traction free boundary condition in the physical, deformed configuration}. This boundary condition is therefore equivalent to the case considered in \cite{parnell2012nonlinear}.

%The elastic wave equations of the hyperelastic cloaks with pre-deformation are obtained from the 'small on large' theory \cite{ogden2007incremental} by deriving the incremental dynamic equations.
\wjp{In the following we employ notation consistent with \cite{ogden2007incremental}. For a given geometry, imposed tractions on the surface of the body in question give rise to its nonlinear deformation and this static deformation is governed by the equilibrium equations
\begin{align}
\textnormal{Div}\mathbf{S} = 0 \label{equil1}
\end{align}
where Div refers to the divergence with respect to $\mathcal{B}_r$ and $\mathbf{S}=\partial W/\partial\mathbf{F}$ (noting that Ogden employs the notation $S_{ij}=\partial W/\partial F_{ji}$) is the Nominal stress tensor, which is related to the (physical) Cauchy stress $\boldsymbol{\sigma}$ via the relation $\mathbf{S}=J\mathbf{F}^{-1}\boldsymbol{\sigma}$, where $J^2=$det$\mathbf{F}$. Stresses are here derived from the strain energy function $W=W(\mathbf{F})$. The imposed traction boundary condition takes the form
\begin{align}
\mathbf{S}^T\mathbf{N} &= \boldsymbol{\tau}(\mathbf{X},\mathbf{F}) \label{BCinc}
\end{align}
with respect to the boundary $\partial\mathcal{B}_r$, having outward pointing normal $\mathbf{N}$ for some imposed function $\boldsymbol{\tau}(\mathbf{X},\mathbf{F})$. It is noted that the problem is most conveniently formulated with respect to $\mathcal{B}_r$ since one knows the geometry of $\mathcal{B}_r$ and therefore one knows $\mathbf{N}$.
}

\wjp{Suppose now that the medium is deformed slightly from the configuration $\mathcal{B}_r$ to say $\dot{\mathcal{B}}_r$ with associated coordinates $\dot{\mathbf{x}}$ via some small displacements $\mathbf{u}=\dot{\mathbf{x}}-\mathbf{x}$. Such a linear elastodynamic response in a general compressible hyperelastic material with pre-deformation is governed by the following equation, derived from the theory of  'small on large' \cite{ogden2007incremental}}
\begin{equation}\label{wave_eq}
\frac{\partial}{\partial X_j} \Big( \mathbb{A}_{ijk\ell}\frac{\partial u_k}{\partial X_\ell} \Big) = \rho_0 \frac{\partial^2 u_i}{\partial t^2},
\end{equation}
%\pz{
%\begin{equation}\label{wave_eq}
%\frac{\partial}{\partial X_j} \Big( \mathbb{A}_{ijk\ell}\frac{\partial u_k}{\partial X_\ell} \Big) - \frac{\partial \dot p}{\partial X_i} = \rho_0 \frac{\partial^2 u_i}{\partial t^2},
%\end{equation}
%}
where $u_i$ are components of the displacement vector $\mathbf{u}$, %\pz{$\dot{p}$ is the incremental penalty pressure,}
$\rho_0$ is the mass density in the undeformed state, $t$ is time, and $\mathbb{A}_{ijk\ell} = \partial^2 W/\partial F_{ij} \partial F_{k\ell}$ is the elasticity tensor dependent on the pre-deformation field $\mathbf{F}$ once the strain energy function $W(\mathbf{F})$ is provided. Therefore the wave equation for the cloaked problem is obtained by setting the elasticity tensor $\mathbb{A}=\mathbb{A}(\mathbf{F})$ in Eq. \eqref{wave_eq} where $\mathbf{F}$ depends on local pre-deformation. On the other hand, the wave equation of the uncloaked problem is obtained by taking $\mathbb{A}=\mathbb{A}(\mathbf{I})$ as the elasticity tensor without pre-deformation, where $\mathbf{I}$ is the identity tensor. An exact hyperelastic cloak is achieved if and only if the wave equation \eqref{wave_eq} is invariant under pre-deformation. In other words, the elasticity tensor $\mathbb{A}$ must be independent of the deformation gradient $\mathbf{F}$, i.e.
\begin{equation} \label{condition}
	\mathbb{A}(\mathbf{F}) = \mathbb{A}(\mathbf{I})
\end{equation}
and as described above the incremental boundary condition in the deformed configuration must also remain invariant to ensure full elastodynamic invariance to pre-deformation. \wjp{The general incremental boundary condition is written most conveniently with respect to the undeformed configuration, as (see (3.10) of \cite{ogden2007incremental})
\begin{align}
\dot{\mathbf{S}}^T\mathbf{N}=\dot{\boldsymbol{\tau}} \label{incBC}
\end{align}
where $\dot{\mathbf{S}}=\mathbb{A}\dot{\mathbf{F}}$ with $\dot{\mathbf{F}}=\partial\dot{\mathbf{x}}/\partial\mathbf{X}$ and $\dot{\boldsymbol{\tau}}$ is the perturbation to the inhomogeneous part of the boundary condition \eqref{BCinc}.}

The necessary condition \eqref{condition} for an exact hyperelastic cloak is so strong that very few strain energy functions are suitable, e.g. the neo-Hookean solid for antiplane waves and semi-linear materials for in-plane waves \cite{norris2012hyperelastic, zhang2017soft}.

\wjp{Here we consider the restriction to incompressible materials and antiplane waves. The imposed traction is planar in the $X_1 X_2$ plane, with the associated condition that the static antiplane displacement is zero. The standard approach for dealing with the constraint of incompressibility is to introduce the Lagrange multiplier $p$, which modifies the nominal stress to be $\mathbf{S}=\partial W/\partial \mathbf{F}-p\mathbf{F}^{-1}$ and in general $p\neq 0$ but in the deformation here $p=p(x_1,x_2)$ since the deformation is purely planar. The incremental nominal stress takes the form
\begin{align}
\dot{\mathbf{S}} = \mathbb{A}\dot{\mathbf{F}} - \dot{p}\mathbf{F}^{-1} + p\mathbf{F}^{-1}\dot{\mathbf{F}}\mathbf{F}^{-1} \label{incnom}
\end{align}
where $\dot{p}$ is the increment to the Lagrange multiplier associated with the incremental perturbation from equilibrium. The incremental equations \eqref{wave_eq} become
\begin{equation}\label{wave_eq2}
\frac{\partial}{\partial X_j} \Big( \mathbb{A}_{ijk\ell}\frac{\partial u_k}{\partial X_\ell} \Big) - \frac{\partial \dot p}{\partial X_i} = \rho_0 \frac{\partial^2 u_i}{\partial t^2}.
\end{equation}
Since the incremental deformation is antiplane motion only, $\dot{p}=0$ \cite{parnell2012nonlinear} and furthermore there is no contribution from $p$ to the antiplane components of the incremental nominal stress defined in \eqref{incnom}. Furthermore, since we anticipate that the forcing is due to the a distant antiplane source in the upper half-space, there is no additional incremental traction condition at the boundary so that $\dot{\boldsymbol{\tau}}=0$ in \eqref{incBC}. All of the above means that the antiplane wave equation takes the form
 \begin{equation}\label{wave_eq3}
\frac{\partial}{\partial X_j} \Big( \mathbb{A}_{3j3\ell}\frac{\partial u_3}{\partial X_\ell} \Big)  = \rho_0 \frac{\partial^2 u_3}{\partial t^2},
\end{equation}
and since $\mathbf{N}=-\mathbf{E}_2$, where $\mathbf{E}_j$ is the unit basis vector pointing in the $X_j$ direction, the incremental boundary condition is simply $\dot{S}_{23}N_2=0$ or rather
\begin{align}
\mathbb{A}_{2323}\dot{F}_{32}+\mathbb{A}_{2313}\dot{F}_{31} &= 0.
\end{align}
This is equivalent to the condition of the incremental antiplane boundary condition being traction free \cite{parnell2012nonlinear}. It should be noted that the incremental in-plane condition would be more complex and certainly the initial deformation \textit{would} influence this boundary condition, unlike the antiplane case.
}

Since the key issue in designing hyperelastic cloaks is to find strain energy functions that satisfy the condition \eqref{condition}, the elasticity tensors of three popular hyperelastic materials are derived, i.e.\ those arising from neo-Hookean, Arruda-Boyce, and Mooney-Rivlin materials. We then assess their suitability to be deployed as hyperelastic ground cloaks.
%Throughout we note that since the boundary condition imposing the local deformation on the surface of the half-space (creating the cloaked region) is in-plane only and therefore we assume that it does not affect the subsequent antiplane elastic wave in the deformed configuration.

(1) \textbf{Neo-Hookean}. The strain energy function for the neo-Hookean solid is given by \cite{holzapfel2000nonlinear}
\begin{equation} \label{W_NH}
	W^{\mathrm{NH}} = \mu (I_1-3)/2,
\end{equation}
where $\mu$ is the shear modulus, $I_1 = \mathrm{tr} \mathbf{C}$ is the first invariant of the Cauchy-Green tensor $\mathbf{C} = \mathbf{F}^T \mathbf{F}$. The elasticity tensor $\mathbb{A}$ is derived as
\begin{equation} \label{A_NH}
	\mathbb{A}^{\mathrm {NH}}_{ijk\ell} = \mu \delta_{ik} \delta_{j\ell},
\end{equation}
where $\delta_{ij}$ is the Kronecker delta tensor. It is noted from Eq. \eqref{A_NH} that the elasticity tensor $\mathbb{A}^{\mathrm{NH}}$ is independent of $\mathbf{F}$ so that the condition \eqref{condition} is always satisfied. This is the reason why a neo-Hookean solid can be used as an exact antiplane cloak \cite{parnell2012nonlinear,parnell2012employing}.

(2) \textbf{Arruda-Boyce}. The Arruda-Boyce material \cite{arruda1993three} is a physics-based model inspired by microstructural modelling and considers the limit stretch of polymer chains and the Langevin statistics. It has proved to be a relatively realistic model with good agreement with experimental data. The strain energy function for the Arruda-Boyce model, when considered as a five term approximation is
\begin{equation} \label{W_AB}
    W^{\mathrm{AB}} = C_1 \sum_{n=1}^{5} \alpha_n \beta^{n-1}(I_1^n-3^n),
\end{equation}
where $C_1$ and $\beta=\lambda_m^{-2}$ are material constants, $\lambda_m$ is the limit stretch of polymer chains, and $\alpha_n \ (n=1,2,3,4,5)$ are coefficients given in \cite{arruda1993three}. The Arruda-Boyce model degenerates to the neo-Hookean solid when the limit stretch $\lambda_m \rightarrow \infty $. A key similarity between the neo-Hookean solid and the Arruda-Boyce model is that both depend on the invariant $I_1$ only. The elasticity tensor of the Arruda-Boyce model is derived, after tedious procedures, as
\begin{equation} \label{A_AB}
	\mathbb{A}_{ijk\ell}^{\mathrm{AB}} = 2 C_1 \sum_{n=1}^{5} n \alpha_n (\beta I_1)^{n-1} [2(n-1)I_1^{-1} F_{ij} F_{k\ell} + \delta_{ik} \delta_{j\ell}].
\end{equation}
The material constant $C_1$ can be expressed in terms of the shear modulus $\mu$ by using the consistency condition. Given the fact that the initial shear modulus of \eqref{A_AB} is equal to $\mu$ of the neo-Hookean solid, it yields
\begin{equation} \label{consistence_AB}
	2C_1 \sum_{n=1}^{5}n \alpha_n (2\beta)^{n-1} = \mu.
\end{equation}
Hence the material constant $C_1$ can be determined from \eqref{consistence_AB} once the initial shear modulus $\mu$ and limit stretch $\lambda_m$ are prescribed. It can be found from Eq. \eqref{A_AB} that the elasticity tensor $\mathbb{A}^{\mathrm{AB}}$ depends on the deformation gradient $\mathbf{F}$. Therefore, the Arruda-Boyce material cannot be used to design an \textit{exact} hyperelastic cloak. However, it will be shown in \S \ref{sec:results} that it is suitable for an approximate cloak. Moreover, it is easy to verify that Eq. \eqref{A_AB} will degenerate to Eq. \eqref{A_NH} if the terms with $n>1$ are all omitted. \pz{Another frequently used model is the Gent material model, which employs a logarithmic strain energy function \cite{gent1996new}. However, it has been shown that the Gent model behaves very similarly to  the Arruda-Boyce model \cite{boyce1996direct} and therefore we do not consider this case further here.}

(3) \textbf{Mooney-Rivlin}. The Mooney-Rivlin model \cite{holzapfel2000nonlinear} is also a generalization of the neo-Hookean solid. Different from the Arruda-Boyce model, the strain energy function of the Mooney-Rivlin model also depends on the second invariant $I_2 = (I_1^2-\mathrm{tr}\mathbf{C}^2)/2$, as
\begin{equation} \label{W_MR}
	W^{\mathrm{MR}} = C_{10}(I_1-3)+C_{01}(I_2-3),
\end{equation}
where $C_{10}$ and $C_{01}$ are two material constants. The elasticity tensor corresponding to \eqref{W_MR} is
\begin{equation} \label{A_MR}
	\mathbb{A}^{\mathrm{MR}} = 2 C_{10} \delta_{ik} \delta_{j\ell} + 2 C_{01}[(I_1 \delta_{j\ell}-C_{j\ell})\delta_{ik}+2F_{ij}F_{k\ell}-F_{i\ell}F_{kj}-F_{im}F_{km}\delta_{jl}].
\end{equation}
On the other hand, the consistency condition gives
\begin{equation} \label{consistence_MR}
	2(C_{10}+C_{01}) = \mu.
\end{equation}
It is clear from Eq. \eqref{W_MR} that the Mooney-Rivlin model deviates from the neo-Hookean solid as the ratio $C_{01}/C_{10}$ increases. As a result, the elasticity tensor $\mathbb{A}^{\mathrm{MR}}$ in \eqref{A_MR} contains a constant term and term that is $\mathbf{F}$-dependent. In theory then, it is impossible to use the Mooney-Rivlin solid to achieve exact cloaking effects but an approximate cloak is possible when the ratio $C_{01}/C_{10}$ is small enough, as should be expected since this is the neo-Hookean limit, as will be shown in \S \ref{sec:results}.

%For the antiplane wave propagation problem considered in this paper, the wave displacement components $u_1$ and $u_2$ are set to zero, and the derivative $\partial/\partial X_3$ also vanishes in Eq. \eqref{wave_eq}. Thus, one can obtain the antiplane wave equation by simply taking $i=k=3$ and $j,\ell = 1,2$ in \eqref{wave_eq}.

\section{Simulation Model}\label{sec:model}

The cloaked region is created by imposing local deformation on the surface, near the origin of the half-space. This problem cannot be solved analytically since this initial deformation of the surface involves a nonlinear, large deformation of the half-space with resulting stress field being localized near the deformation and decaying away from this region. The deformation is inhomogeneous and is a combination of large stretch and rotation. Hence, finite element simulation is performed by using the commercial software ABAQUS 6.14-3 \cite{abaqusManual}. Two steps are conducted for the cloaking simulation including a static contact step generating the cloaked region and an incremental wave propagation step to obtain the wave field. In contrast, only the wave propagation step is needed for the uncloaked situation. For the contact step, the static antiplane displacement is set to zero. However, the in-plane displacements $u_1$ and $u_2$ are fixed as zero for the incremental deformation, assuming that only antiplane wave propagation prevails. In all simulation examples, the bottom surface of the bulk region is kept antiplane traction free and frictionless. Both the static contact and incremental wave steps are solved by using the  explicit dynamics solver in ABAQUS. Note that the mass scaling has been used for the static step to improve numerical efficiency.

\begin{figure}[htbp!]
	\centering
	\includegraphics[width=9cm]{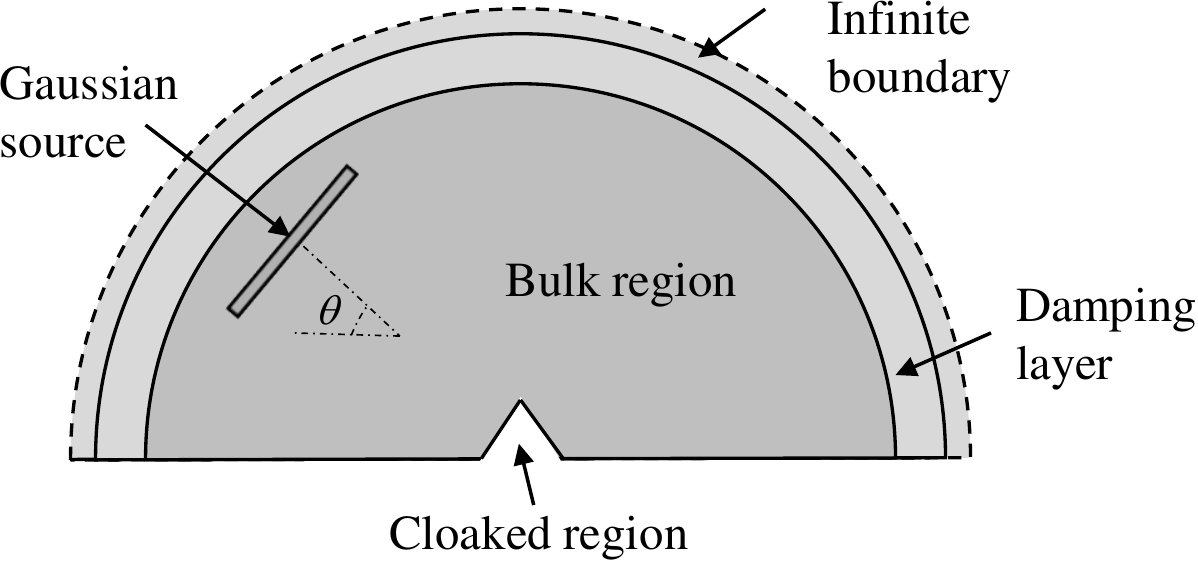}
	\caption{\label{fig:model} Schematic illustration of the simulation model for a hyperelastic ground cloak.}
\end{figure}
The geometric model, \wjp{consisting of five distinct regions}, is illustrated in Fig.\ \ref{fig:model}. The major part is the bulk region with a radius of 0.2 m. The cloaked region is created in the static contact step by using a rigid smooth triangular wedge of $120^{\circ}$ with \pz{a fillet radius of 0.01 m to avoid the stress singularity that would otherwise arise for the sharp end. The triangular wedge is modeled as a rigid surface in ABAQUS and a rigid body displacement loading is applied on the wedge to create an indentation depth of 0.02 m.}  A Gaussian wave source is embedded in the bulk region to generate a Gaussian beam that propagates towards the indentation that forms the ground cloak. The propagation direction of the beam makes an angle of \pz{ $\theta$} with the undeformed surface of the half-space. The frequency of the Gaussian beam is $f=20$ KHz and its beam waist is $w=2 \lambda$ with $\lambda$ indicating the shear wavelength. Adjacent to the bulk region are the damping layer and infinite boundary layer introduced in order to absorb the outgoing waves \cite{liu2003non, rajagopal2012use}. Note that the damping layer is required since the infinite boundary layer alone is not effective enough to absorb waves in ABAQUS, especially for the oblique incidence situation. The damping layer has a total thickness of 0.02 m and is further divided into $N$ sub-layers. Rayleigh damping is used with a mass proportional damping factor $c$ that increases gradually as \cite{rajagopal2012use}
\begin{equation}\label{damping}
	c_i = c_0 (i/N)^3, \quad (i=1,2...,N),
\end{equation}
where $i$ indicates the sub-layer number, $c_0=4\pi f$ is the maximum damping factor, and $N$ is taken as 10 in this work, indicating 10 sub-layers.

The whole model in Fig.\ \ref{fig:model} consists of only one layer of elements along the thickness direction since the antiplane wave problem is considered in this work. In addition, the two layers of nodes on the front and back surfaces are coupled together. The eight-node three-dimensional element C3D8R with reduced integral is used for the whole domain except the infinite boundary, in which the infinite element CIN3D8 is chosen. The element size is approximately equal to $\lambda/16$ to achieve both accuracy and efficiency. \wjp{It has been verified that such a mesh size will result in converged static and incremental wave solutions.}

\wjp{The material models are chosen from the pre-defined neo-Hookean, Arruda-Boyce, and Mooney-Rivlin hyperelastic material strain energy functions in ABAQUS 6.14-3 \cite{abaqusManual}}. The initial shear modulus is taken as $\mu=100$ MPa and the Poisson's ratio is 0.495 indicating that the medium is almost incompressible, as required. In addition, the mass density is $\rho_0=1000 \ \mathrm{kg}\mathrm{m}^{-3} $ for all materials. As a result, the initial shear wave length is $\lambda=f^{-1}\sqrt{\mu/\rho_0}=1.58 \times 10^{-2} $ m. The total simulation time for the wave step is $t=$ 1.45 ms.  For the Arruda-Boyce material, the limit stretch parameter $\lambda_m$ is tuned to investigate its effect on the wave scattering in a hyperelastic cloak, while the influence of the parameter $C_{01}/C_{10}$ is explored for the Mooney-Rivlin cases.

\section{Results}\label{sec:results}

The wave scattering behaviour in uncloaked configurations and configurations associated with hyperelastic cloaks is discussed in this section. An antiplane Gaussian beam is generated by the Gaussian source and the wave amplitude fields $|u_3|$ are obtained and discussed. All wave field snapshots are taken at  $t=$1.45 ms of the wave step of the simulation.

\begin{figure}[htbp!]
	\centering
	\includegraphics[width=16cm]{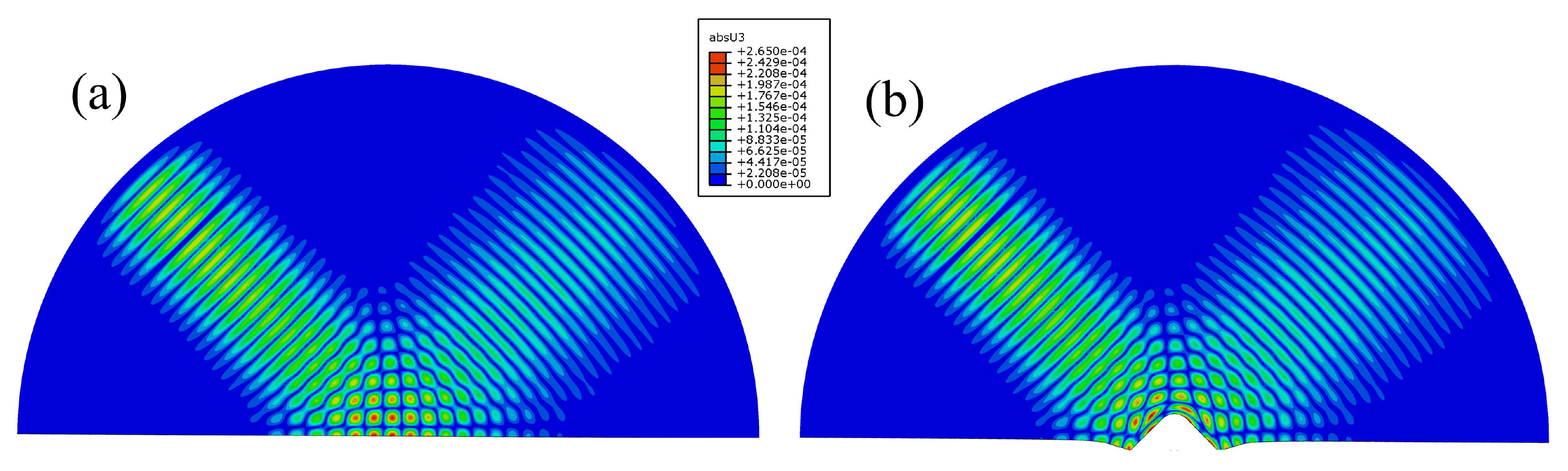}
	\caption{\label{fig:NHfield} Antiplane wave fields for hyperelastic ground cloaks using neo-Hookean solids \pz{($\theta=45^{\circ}$)}. (a) Scattering from the surface in an undeformed medium. (b) Cloaked case created by imposing pre-deformation at the bottom center. The wave field is invariant after pre-deformation for a neo-Hookean type cloak, although the wave field near the cloaked region is distorted due to the mapping. }
\end{figure}
Fig.\ \ref{fig:NHfield} shows the propagation and scattering of the antiplane waves in a neo-Hookean cloak before and after imposing pre-deformation \pz{when the incident angle is $\theta=45^{\circ} $}. Since the cloaked region is created by local pre-deformation, it is clear that Fig. \ref{fig:NHfield} (a) shows the uncloaked case, while Fig. \ref{fig:NHfield} (b) illustrates the cloaked case. The Gaussian beam is incident from the upper left corner and reflected to the upper right direction. It is observed by comparing Figs. \ref{fig:NHfield} (a) and (b) that the pre-deformation does not change the scattered wave field in a neo-Hookean type cloak, which is consistent with the cylindrical cloak described the literature \cite{parnell2012nonlinear, parnell2012employing}. A quantitative metric is used to compare the wave fields for the uncloaked and cloaked scenarios. If we denote the wave field as two vectors $\mathbf{w}^u = \{ u_{3i}^u \}$ and $\mathbf{w}^c = \{ u_{3i}^c \}$ in the uncloaked and cloaked cases, respectively, where $i$ represents all nodes and the superscript $u$ or $c$ indicates the uncloaked or cloaked field. The similarity between these two fields is defined as
\begin{equation} \label{similarity}
	\mathrm{similarity} = \frac{\mathbf{w}^u \cdot \mathbf{w}^c}{\lVert \mathbf{w}^u \rVert \ \lVert \mathbf{w}^c \rVert}.
\end{equation}
The similarity \eqref{similarity} is also called the cosine similarity between two fields, which is equal to 1 for two identical fields and $-1$ for two opposite fields. According to Eq. \eqref{similarity}, it is calculated that the similarity between the two wave fields in Fig. \ref{fig:NHfield} is 98.2 $ \% $, slightly lower than the theoretical value as 100 $\%$. This marginal error is almost certainly induced by the numerical implementation in the software ABAQUS to map stress quantities between different configurations, to control element distortion and volume locking, etc. But it will also be due to the fact that we employ an almost-incompressible medium here whereas the exact invariance is associated with the perfect incompressible strain energy function. To conclude, the finite element analysis confirms that the almost incompressible neo-Hookean solid is able to achieve an almost-perfect cloaking effect for antiplane waves.

\begin{figure}[htbp!]
	\centering
	\includegraphics[width=16cm]{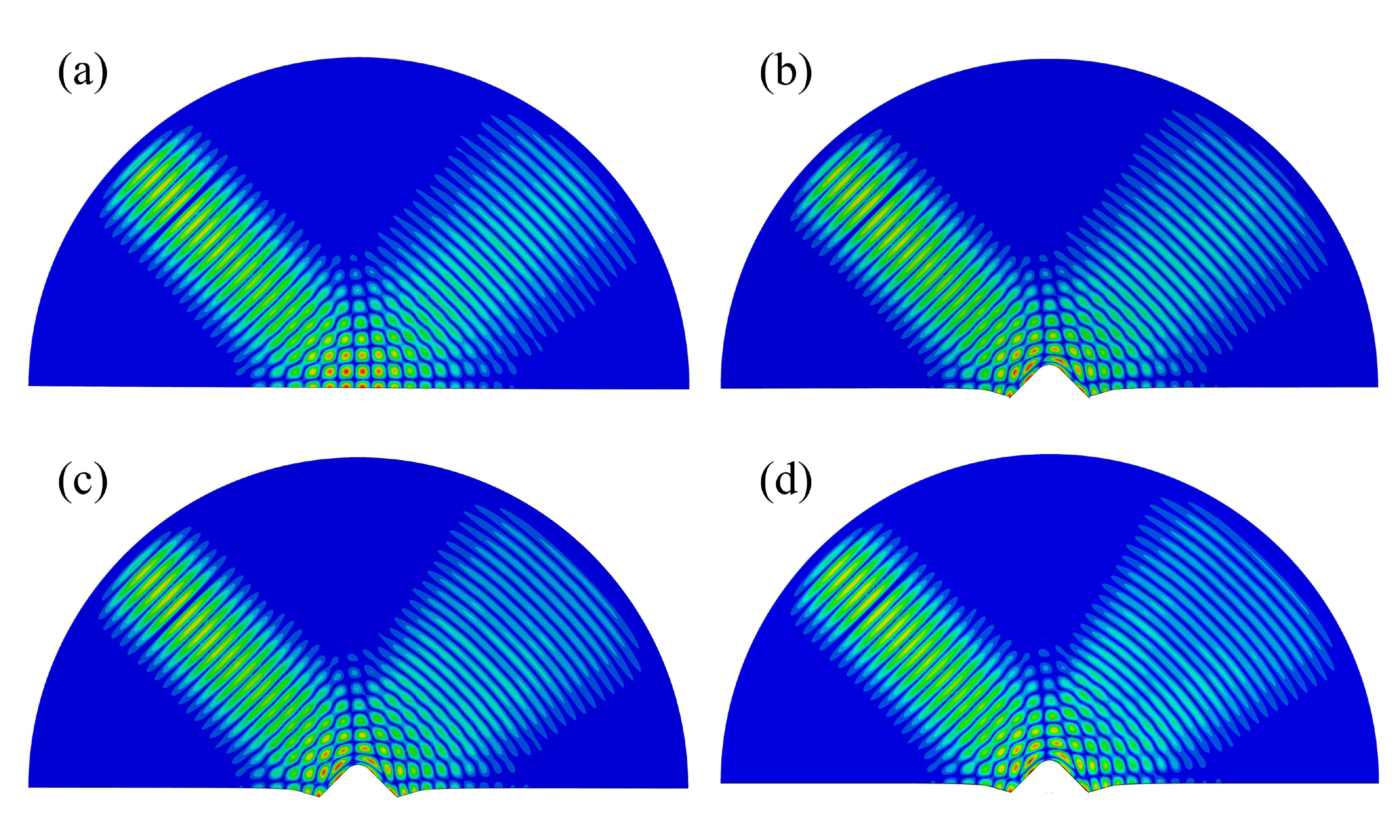}
	\caption{\label{fig:ABfield} Antiplane wave fields for hyperelastic ground cloaks using Arruda-Boyce materials \pz{($\theta=45^{\circ}$)}. (a) Scattering from the surface in an undeformed medium. (b) Cloaked case with $\lambda_m = 7$. (c) Cloaked case with $\lambda_m=3$. (d) Cloaked case with $\lambda_m=2.5$. The Arruda-Boyce material deviates from the neo-Hookean solid as the limit stretch $\lambda_m$ decreases.}
\end{figure}
\begin{figure}[htbp!]
	\centering
	\includegraphics[width=9cm]{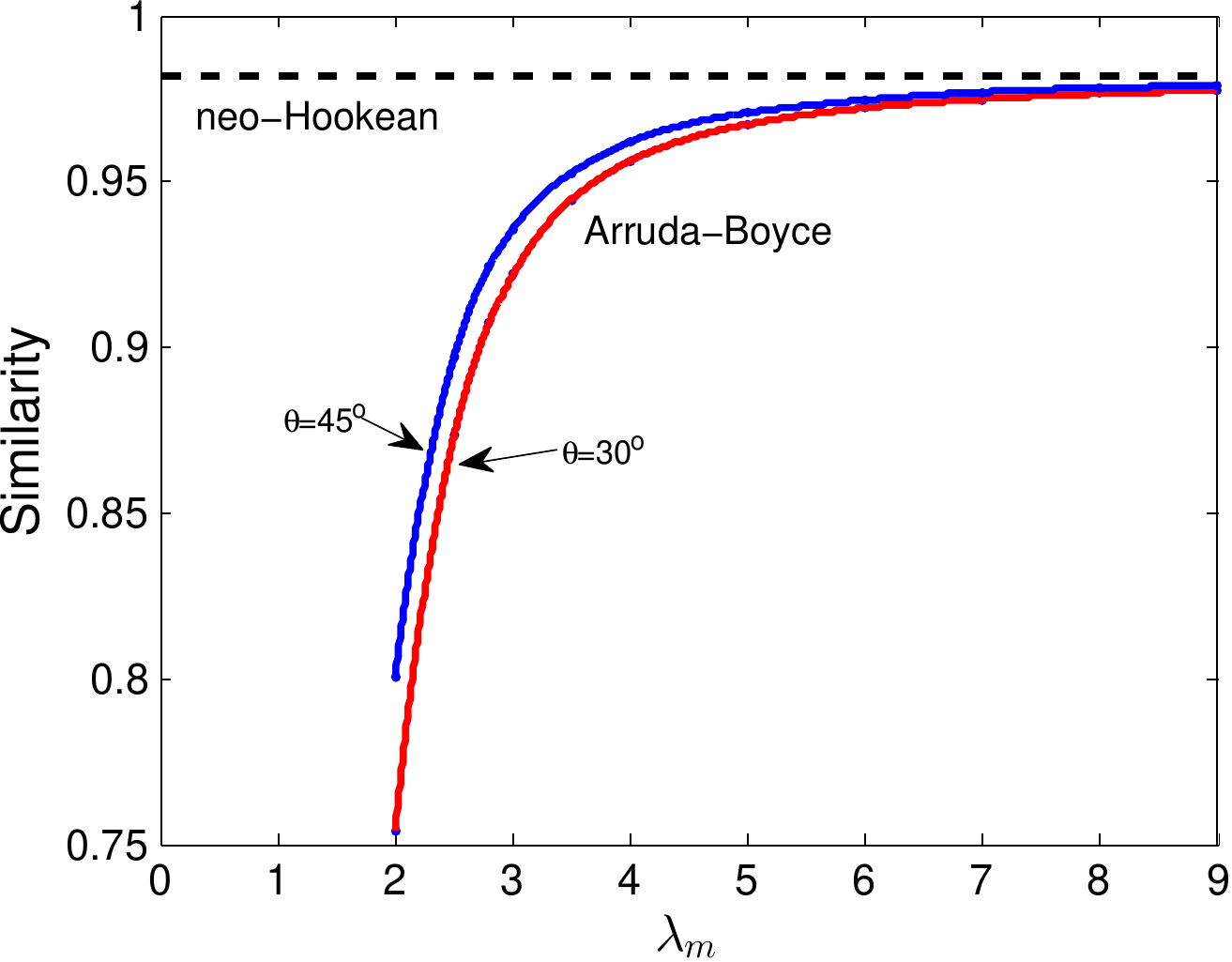}
	\caption{\label{fig:AB} Comparison between the wave scattering in uncloaked and cloaked cases in hyperelastic ground cloaks using Arruda-Boyce materials. Approximate cloaking is achieved when the limit stretch $ \lambda_m>3 $.}
\end{figure}

The antiplane wave response in hyperelastic cloaks consisting of Arruda-Boyce materials are illustrated in Fig.\ \ref{fig:ABfield}. The color scale is the same as that in Fig.\ \ref{fig:NHfield}.  Figure\ \ref{fig:ABfield} (a) illustrates the wave scattering in a hyperelastic cloak medium without any pre-deformation. This solution is identical to that in Fig.\ \ref{fig:NHfield} (a) since the initial shear modulus of the Arruda-Boyce materials \pz{are} the same as that of neo-Hookean according to the consistency condition \eqref{consistence_AB}. The wave fields in pre-deformed cloaks are illustrated in Figs.\ \ref{fig:ABfield} (b), (c) and (d) for different limit stretch parameters $\lambda_m =$ 7, 3 and 2.5, respectively. It is observed that the scattered wave fields in Figs.\ \ref{fig:ABfield} (b), (c) and (d) deviate from the undeformed case in Fig.\ \ref{fig:ABfield} (a) more and more as $\lambda_m$ decreases. This is because the influence of the higher order terms ($n>1$) in the elasticity tensor $\mathbb{A}^{\mathrm{AB}}$ in \eqref{A_AB} play a more significant role when the limit stretch $\lambda_m$ decreases so that the elasticity tensor becomes ever more deformation-dependent. Overall, the scattered wave patterns for the cloaked cases are not significantly different from the scattering in the initial configuration although the reflected beam width becomes wider when $\lambda_m$ decreases.

Further, quantitative analysis of the comparison between the wave fields in the undeformed and deformed/cloaked cases is shown in Fig.\ \ref{fig:AB} with the similarity defined as per \eqref{similarity}. It is found that the similarity for the cloaked cases in Fig.\ \ref{fig:ABfield} (b), (c) and (d) are $97.7 \%$, $93.6 \%$ and $89.7 \%$, respectively, when they are compared to the undeformed case. Figure\ \ref{fig:AB} shows that the Arruda-Boyce materials behave almost the same as the neo-Hookean solid when $\lambda_m>5$, and very high similarity is achieved when $\lambda_m>3.5$. Although the similarity in Fig.\ \ref{fig:AB} decreases rapidly when $\lambda_m<3$, most realistic rubbers have the limit stretch $\lambda_m$ greater than 3 \cite{arruda1993three}. \pz{Figure \ \ref{fig:AB} also shows the similarity when the incident angle is $\theta=30^{\circ}$, which indicates that the incident angle will slightly affect the cloaking response. The corresponding wave fields are illustrated in the Supplementary Material.} The results indicate that Arruda-Boyce materials are excellent candidates for approximate cloaks.

\begin{figure}[htbp!]
	\centering
	\includegraphics[width=16cm]{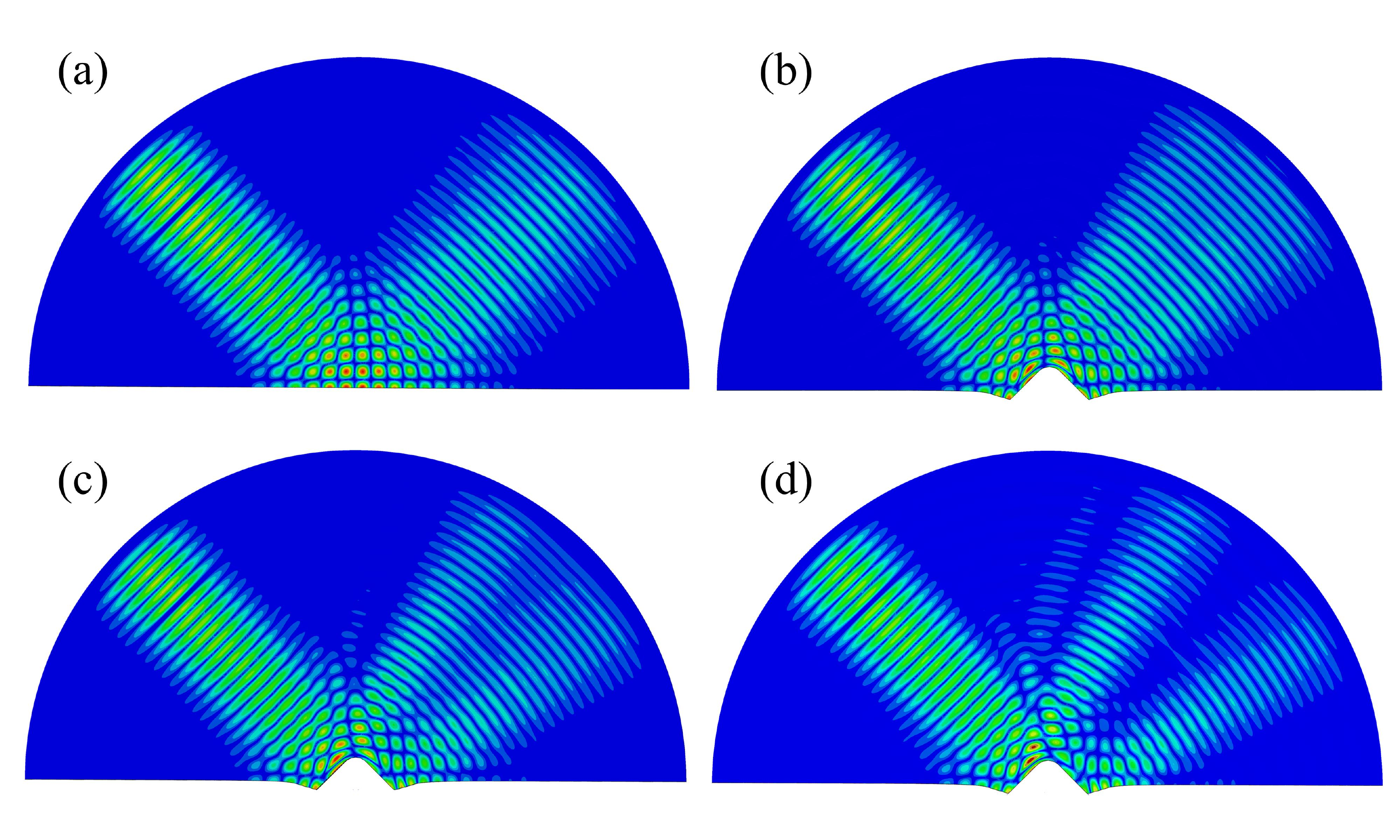}
	\caption{\label{fig:MRfield} Antiplane wave fields for hyperelastic ground cloaks using Mooney-Rivlin materials \pz{($\theta=45^{\circ}$)}. (a) Scattering from the surface in an undeformed medium. (b) Cloaked case with $C_{01}/C_{10} = 0.03$. (c) Cloaked case with $C_{01}/C_{10} = 0.07$. (d) Cloaked case with $C_{01}/C_{10} = 0.15$. }
\end{figure}
\begin{figure}[htbp!]
	\centering
	\includegraphics[width=9cm]{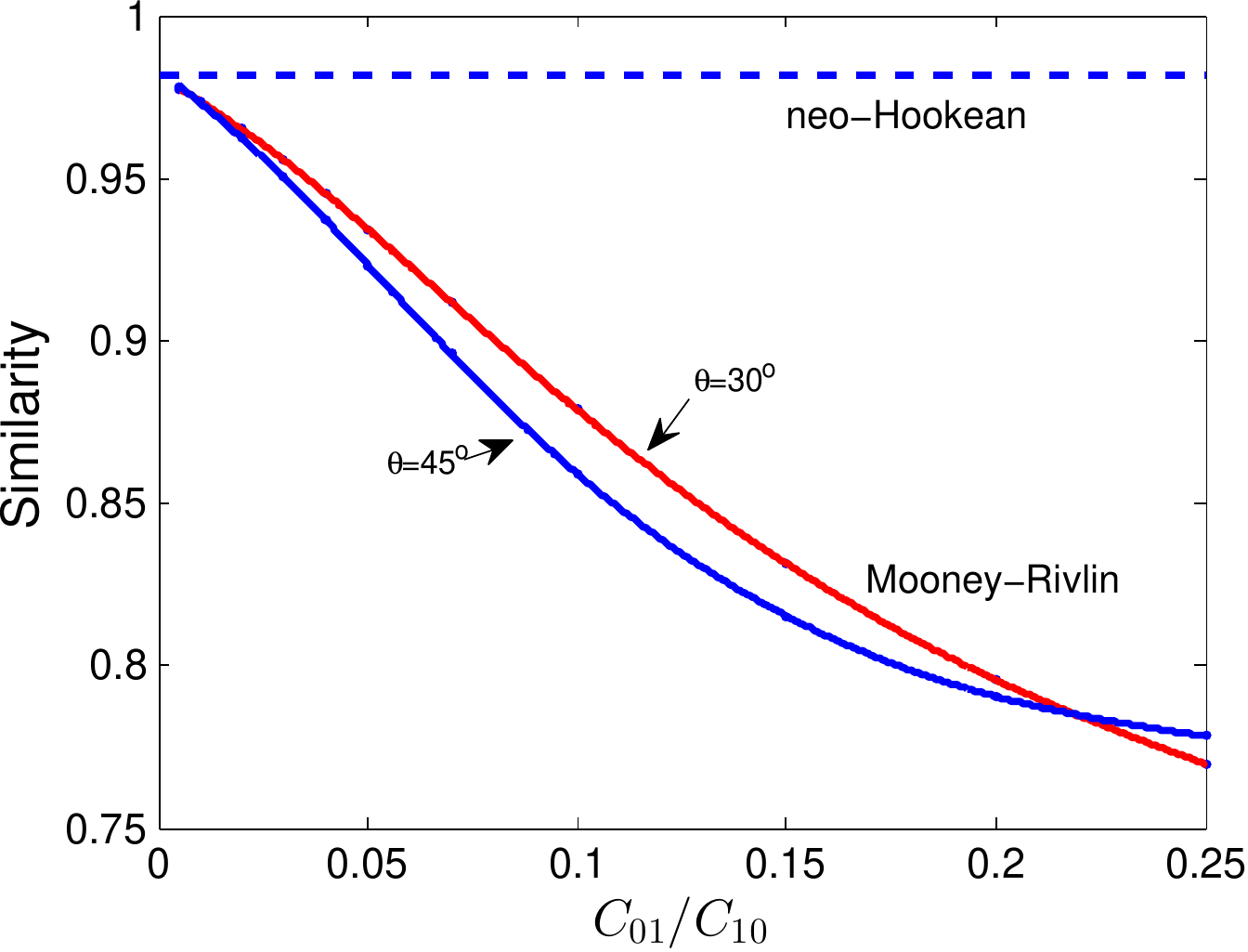}
	\caption{\label{fig:MR} Comparison between the wave scattering in uncloaked and cloaked cases in hyperelastic ground cloaks using Mooney-Rivlin materials. The Mooney-Rivlin materials can be used as approximate cloaks only when the ratio $C_{01}/C_{10}<0.03$.}
\end{figure}
It has been shown in Figs. \ref{fig:NHfield} and \ref{fig:ABfield} that the neo-Hookean and Arruda-Boyce materials can be used to design exact or approximate cloaks. A shared feature of these two materials is that their strain energy functions \eqref{W_NH} and \eqref{W_AB} are dependent only on the first invariant $I_1$. Therefore, it would be interesting to explore the effect of the second invariant $I_2$ on the antiplane wave responses in a hyperelastic cloak. To achieve this aim, we consider hyperelastic cloaks using the Mooney-Rivlin material with strain energy defined in \eqref{W_MR}. The initial shear modulus of the Mooney-Rivlin materials is kept as $\mu$, the same as the neo-Hookean and Arruda-Boyce materials. The ratio between the two material coefficients $C_{01}/C_{10}$ is studied since this ratio is a measure of the similarity between the Mooney-Rivlin material and the neo-Hookean solid.

Figure\ \ref{fig:MRfield} shows the antiplane wave fields in Mooney-Rivlin type cloaks \pz{($\theta=45^{\circ}$)} with color scale the same as that in Fig. \ref{fig:NHfield}. Therein, Fig. \ref{fig:MRfield} (a) illustrates the undeformed case and Figs.\ \ref{fig:MRfield} (b), (c) and (d) show the deformed scenarios with different $C_{01}/C_{10}$ values. By comparing Fig. \ref{fig:MRfield} (a) with (b), it is found that the reflected beam in the cloaked case is not much different from that in the undeformed situation since the ratio $C_{01}/C_{10}$ is small enough ($<0.03$). However, when the ratio $C_{01}/C_{10}$ increases, the reflected beam is split into two branches, as what is shown in Fig. \ref{fig:MRfield} (c) and (d). The cause of this phenomenon is that the $C_{01}$ -dependent term in the elasticity tensor $\mathbb{A}^{\mathrm{MR}}$ in \eqref{A_MR} is very sensitive to the magnitude of the coefficient. Figure\ \ref{fig:MR} shows the similarity between the wave fields for undeformed and deformed cases when the Mooney-Rivlin material is adopted. The similarity values corresponding to Figs. \ref{fig:MRfield} (b), (c) and (d) are 95.0\%, 89.6\% and 81.5\%, respectively. It is found that the similarity of a Mooney-Rivlin cloak deviates rapidly from the neo-Hookean type as the ratio $C_{01}/C_{10}$ increases.  However, the Mooney-Rivlin cloak can achieve relatively high similarity when the ratio $C_{01}/C_{10}<0.03$, implying that an approximate cloak is still possible, although the material property range is very limited. One expects that this influence is associated with the dependence on the second invariant $I_2$ and therefore when considering practical realisations of hyperelastic cloaks this should be considered carefully. \pz{The wave scattering responses are similar when the incident angles of the beams are changed to $\theta=30^{\circ}$ (see Supplementary Material). The corresponding similarity factors are shown in Fig. \ref{fig:MR}, which verifies the robust performance of the hyperelastic cloaks.}

\section{Conclusions}\label{sec:conclusion}

Hyperelastic cloaking offers a convenient way to design elastic cloaks in lieu of employing microstructured materials. \wjp{In particular they offer the potential advantages of frequency independence, tunability and unlike their microstructured counterparts they are not fixed once in-situ. Moreover the cloaked region can be made larger or smaller by imposing more or less deformation.} The key issue in designing hyperelastic cloaks is to find suitable materials with a deformation-independent elasticity tensor to guarantee the form-invariance of the wave equations. Unfortunately, existing research suggests that an exact hyperelastic cloak can only be achieved for special hyperelastic materials like the neo-Hookean solid and semi-linear materials. This work has explored the possibility of using other hyperelastic materials to design antiplane ground cloaks in order to expand the material selection space. Two representative hyperelastic materials are considered: an $I_1$ -dependent material (Arruda-Boyce) and an $I_2$ -dependent material (Mooney-Rivlin). The hyperelastic cloaks using different material types are simulated by dynamic finite element analysis and evaluated to compare the similarity between the wave fields in the deformed (cloaked) and undeformed configurations. The simulation results confirm that the neo-Hookean solid is able to create an exact antiplane cloak. The Arruda-Boyce material behaves closely to the neo-Hookean solid for a wide range of limit stretch parameters $\lambda_m>3.5$, and in these cases, approximate antiplane cloaks are possible. In contrast, the wave response in the Mooney-Rivlin material is very sensitive to the $C_{01}$ -dependent term, and approximate cloaks are possible only when the ratio $C_{01}/C_{10}$ is small enough ($<0.03$), i.e.\ when it behaves almost like a neo-Hookean medium. Future research can be focused towards evaluating other types of strain energy functions as well as conducting experimental verification for the hyperelastic cloaks. Consideration of the two-dimensional in-plane and three-dimensional elastodynamic scenarios are also of great interest.

\section*{Acknowledgements}
The authors are grateful to the Engineering and Physical Sciences Research Council (EPSRC) for financial support via grant no. EP/L018039/1.

{\small
	
\bibliographystyle{unsrt}
\bibliography{ground_cloak_arxiv}

\begin{thebibliography}{10}

\bibitem{greenleaf2003nonuniqueness}
Allan Greenleaf, Matti Lassas, and Gunther Uhlmann.
\newblock On nonuniqueness for calderon's inverse problem.
\newblock {\em Mathematical Research Letters}, 10(5):685--693, 2003.

\bibitem{pendry2006controlling}
John~B Pendry, David Schurig, and David~R Smith.
\newblock Controlling electromagnetic fields.
\newblock {\em science}, 312(5781):1780--1782, 2006.

\bibitem{leonhardt2006optical}
Ulf Leonhardt.
\newblock Optical conformal mapping.
\newblock {\em Science}, 312(5781):1777--1780, 2006.

\bibitem{ruan2007ideal}
Zhichao Ruan, Min Yan, Curtis~W Neff, and Min Qiu.
\newblock Ideal cylindrical cloak: perfect but sensitive to tiny perturbations.
\newblock {\em Physical Review Letters}, 99(11):113903, 2007.

\bibitem{schurig2006metamaterial}
David Schurig, JJ~Mock, BJ~Justice, Steven~A Cummer, John~B Pendry, AF~Starr,
  and DR~Smith.
\newblock Metamaterial electromagnetic cloak at microwave frequencies.
\newblock {\em Science}, 314(5801):977--980, 2006.

\bibitem{cai2007optical}
Wenshan Cai, Uday~K Chettiar, Alexander~V Kildishev, and Vladimir~M Shalaev.
\newblock Optical cloaking with metamaterials.
\newblock {\em Nature photonics}, 1(4):224--227, 2007.

\bibitem{li2008hiding}
Jensen Li and JB~Pendry.
\newblock Hiding under the carpet: a new strategy for cloaking.
\newblock {\em Physical review letters}, 101(20):203901, 2008.

\bibitem{liu2009broadband}
R~Liu, C~Ji, JJ~Mock, JY~Chin, TJ~Cui, and DR~Smith.
\newblock Broadband ground-plane cloak.
\newblock {\em Science}, 323(5912):366--369, 2009.

\bibitem{ma2010three}
Hui~Feng Ma and Tie~Jun Cui.
\newblock Three-dimensional broadband ground-plane cloak made of metamaterials.
\newblock {\em Nature communications}, 1:21, 2010.

\bibitem{leonhardt2009transformation}
Ulf Leonhardt and Thomas~G Philbin.
\newblock Transformation optics and the geometry of light.
\newblock {\em Progress in Optics}, 53:69--152, 2009.

\bibitem{norris2008acoustic}
Andrew~N Norris.
\newblock Acoustic cloaking theory.
\newblock {\em Proceedings of the Royal Society of London Series A},
  464:2411--2434, 2008.

\bibitem{chen2010acoustic}
Huanyang Chen and Che~Ting Chan.
\newblock Acoustic cloaking and transformation acoustics.
\newblock {\em Journal of Physics D: Applied Physics}, 43(11):113001, 2010.

\bibitem{popa2011experimental}
Bogdan-Ioan Popa, Lucian Zigoneanu, and Steven~A Cummer.
\newblock Experimental acoustic ground cloak in air.
\newblock {\em Physical review letters}, 106(25):253901, 2011.

\bibitem{zigoneanu2014three}
Lucian Zigoneanu, Bogdan-Ioan Popa, and Steven~A Cummer.
\newblock Three-dimensional broadband omnidirectional acoustic ground cloak.
\newblock {\em Nature materials}, 13(4):352--355, 2014.

\bibitem{diatta2014controlling}
Andre Diatta and Sebastien Guenneau.
\newblock Controlling solid elastic waves with spherical cloaks.
\newblock {\em Applied Physics Letters}, 105(2):021901, 2014.

\bibitem{hu2011approximate}
Jin Hu, Zheng Chang, and Gengkai Hu.
\newblock Approximate method for controlling solid elastic waves by
  transformation media.
\newblock {\em Physical Review B}, 84(20):201101, 2011.

\bibitem{liu2015transformation}
Yongquan Liu, Wei Liu, Bing Li, and Xianyue Su.
\newblock Transformation method to control shear horizontal waves.
\newblock {\em International Journal of Applied Mechanics}, 7(03):1550049,
  2015.

\bibitem{chang2012transformation}
Zheng Chang, Xiaoning Liu, Gengkai Hu, and Jin Hu.
\newblock Transformation ray method: Controlling high frequency elastic waves
  (l).
\newblock {\em The Journal of the Acoustical Society of America},
  132(4):2942--2945, 2012.

\bibitem{stenger2012experiments}
Nicolas Stenger, Manfred Wilhelm, and Martin Wegener.
\newblock Experiments on elastic cloaking in thin plates.
\newblock {\em Physical Review Letters}, 108(1):014301, 2012.

\bibitem{climente2016analysis}
Alfonso Climente, Daniel Torrent, and Jos{\'e} S{\'a}nchez-Dehesa.
\newblock Analysis of flexural wave cloaks.
\newblock {\em AIP Advances}, 6(12):121704, 2016.

\bibitem{milton2006cloaking}
Graeme~W Milton, Marc Briane, and John~R Willis.
\newblock On cloaking for elasticity and physical equations with a
  transformation invariant form.
\newblock {\em New Journal of Physics}, 8(10):248, 2006.

\bibitem{norris2011elastic}
Andrew~N Norris and Alexander~L Shuvalov.
\newblock Elastic cloaking theory.
\newblock {\em Wave Motion}, 48(6):525--538, 2011.

\bibitem{brun2009achieving}
Michele Brun, S{\'e}bastien Guenneau, and Alexander~B Movchan.
\newblock Achieving control of in-plane elastic waves.
\newblock {\em Applied physics letters}, 94(6):061903, 2009.

\bibitem{colquitt2014transformation}
Daniel~J Colquitt, Morvan Brun, Massimiliano Gei, Alexander~B Movchan,
  Natasha~V Movchan, and Ian~Samuel Jones.
\newblock Transformation elastodynamics and cloaking for flexural waves.
\newblock {\em Journal of the Mechanics and Physics of Solids}, 72:131--143,
  2014.

\bibitem{sklan2017elastic}
Sophia~R Sklan, Ronald Pak, and Baowen Li.
\newblock Elastic wave cloaking via symmetrized transformation media.
\newblock {\em arXiv preprint arXiv:1709.07926}, 2017.

\bibitem{wu2015manipulation}
Linzhi Wu and Penglin Gao.
\newblock Manipulation of the propagation of out-of-plane shear waves.
\newblock {\em International Journal of Solids and Structures}, 69:383--391,
  2015.

\bibitem{khlopotin2015transformational}
Alexey Khlopotin, Peter Olsson, and Fredrik Larsson.
\newblock Transformational cloaking from seismic surface waves by micropolar
  metamaterials with finite couple stiffness.
\newblock {\em Wave Motion}, 58:53--67, 2015.

\bibitem{colombi2016transformation}
Andrea Colombi, Sebastien Guenneau, Philippe Roux, and Richard~V Craster.
\newblock Transformation seismology: composite soil lenses for steering surface
  elastic rayleigh waves.
\newblock {\em Scientific reports}, 6:25320, 2016.

\bibitem{colombi2017elastic}
Andrea Colombi, Richard Craster, Daniel Colquitt, Younes Achaoui, Sebastien
  Guenneau, Philippe Roux, and Matthieu Rupin.
\newblock Elastic wave control beyond band-gaps: shaping the flow of waves in
  plates and half-spaces.
\newblock {\em arXiv preprint arXiv:1705.09288}, 2017.

\bibitem{norris2012hyperelastic}
Andrew~N Norris and William~J Parnell.
\newblock Hyperelastic cloaking theory: transformation elasticity with
  pre-stressed solids.
\newblock {\em Proceedings of the Royal Society of London A}, 468:2881--2903,
  2012.

\bibitem{ParnellBookChapter2017}
Andrew~N Norris and William~J Parnell.
\newblock Hyperelastic cloaking theory.
\newblock In S.~Maier, K.~Shamonina, S.~Guenneau, O.~Hess, and J.~Aizpurua,
  editors, {\em A handbook of metamaterials and nanophotonics}, volume~2. World
  Scientific, 2017.

\bibitem{parnell2012nonlinear}
William~J Parnell.
\newblock Nonlinear pre-stress for cloaking from antiplane elastic waves.
\newblock {\em Proceedings of the Royal Society of London Series A},
  468:563--580, 2012.

\bibitem{parnell2012employing}
William~J Parnell, Andrew~N Norris, and Tom Shearer.
\newblock Employing pre-stress to generate finite cloaks for antiplane elastic
  waves.
\newblock {\em Applied Physics Letters}, 100(17):171907, 2012.

\bibitem{zhang2017soft}
P~Zhang and WJ~Parnell.
\newblock Soft phononic crystals with deformation-independent band gaps.
\newblock {\em Proceedings of the Royal Society of London A},
  473(2200):20160865, 2017.

\bibitem{parnell2013antiplane}
William~J Parnell and Tom Shearer.
\newblock Antiplane elastic wave cloaking using metamaterials, homogenization
  and hyperelasticity.
\newblock {\em Wave Motion}, 50(7):1140--1152, 2013.

\bibitem{ogden2007incremental}
Ray~W Ogden.
\newblock Incremental statics and dynamics of pre-stressed elastic materials.
\newblock In {\em Waves in nonlinear pre-stressed materials}, pages 1--26.
  Springer, 2007.

\bibitem{holzapfel2000nonlinear}
Gerhard~A Holzapfel.
\newblock {\em Nonlinear solid mechanics}, volume~24.
\newblock Wiley Chichester, 2000.

\bibitem{arruda1993three}
Ellen~M Arruda and Mary~C Boyce.
\newblock A three-dimensional constitutive model for the large stretch behavior
  of rubber elastic materials.
\newblock {\em Journal of the Mechanics and Physics of Solids}, 41(2):389--412,
  1993.

\bibitem{gent1996new}
AN~Gent.
\newblock A new constitutive relation for rubber.
\newblock {\em Rubber chemistry and technology}, 69(1):59--61, 1996.

\bibitem{boyce1996direct}
Mary~C Boyce.
\newblock Direct comparison of the gent and the arruda-boyce constitutive
  models of rubber elasticity.
\newblock {\em Rubber chemistry and technology}, 69(5):781--785, 1996.

\bibitem{abaqusManual}
Simulia.
\newblock {\em Abaqus Analysis User's Guide, Version 6.14}.
\newblock Dassault Systemes, Providence, RI, 2014.

\bibitem{liu2003non}
GR~Liu and SS~Quek Jerry.
\newblock A non-reflecting boundary for analyzing wave propagation using the
  finite element method.
\newblock {\em Finite elements in analysis and design}, 39(5):403--417, 2003.

\bibitem{rajagopal2012use}
Prabhu Rajagopal, Mickael Drozdz, Elizabeth~A Skelton, Michael~JS Lowe, and
  Richard~V Craster.
\newblock On the use of absorbing layers to simulate the propagation of elastic
  waves in unbounded isotropic media using commercially available finite
  element packages.
\newblock {\em NDT \& E International}, 51:30--40, 2012.

\end{thebibliography}

}

\end{document}